\documentclass[aps,preprint,hyperref]{revtex4-1}
\usepackage{graphicx,amsmath}
\begin{document}
\title{Preventing blue sky catastrophe in an array of $N$ identical Van-der-Pol oscillators coupled through nonlinear coupling}
\title{Preventing Blow-ups through Dynamic Random Links}
\title{Taming Blow-ups through Dynamic Random Links}
\title{Taming Explosive Growth through Dynamic Random Links}

\author{Anshul Choudhary}
\affiliation{Indian Institute of Science Education and Research (IISER) Mohali, Knowledge City, SAS Nagar, Sector 81, Manauli PO 140 306, Punjab, India}
\author{Vivek Kohar}
\affiliation{Indian Institute of Science Education and Research (IISER) Mohali, Knowledge City, SAS Nagar, Sector 81, Manauli PO 140 306, Punjab, India}
\author{Sudeshna Sinha\footnote{Correspondence to sudeshna@iisermohali.ac.in}}
\affiliation{Indian Institute of Science Education and Research (IISER) Mohali, Knowledge City, SAS Nagar, Sector 81, Manauli PO 140 306, Punjab, India}

\begin{abstract}
  We study the dynamics of a collection of nonlinearly coupled
  limit cycle oscillators, relevant to systems ranging from neuronal
  populations to electrical circuits, under coupling topologies
  varying from a regular ring to a random network. We find that the
  trajectories of this system escape to infinity under regular
  coupling, for sufficiently strong coupling strengths. However, when
  some fraction of the regular connections are dynamically randomized,
  the unbounded growth is suppressed and the system always remains
  bounded. Further we determine the critical fraction of random links
  necessary for successful prevention of explosive behaviour, for
  different network rewiring time-scales. These results suggest a
  mechanism by which blow-ups may be controlled in extended oscillator
  systems.
\end{abstract}

\maketitle

Collections of coupled dynamical systems were introduced as simple
models capturing the essential features of large interactive systems
\cite{cml}. Research focussed around such distributed systems and
complex networks has provided frameworks for understanding
spatiotemporal patterns in problems ranging from Josephson junction
arrays and multimode lasers, to microfluidic arrays and evolutionary
biology \cite{ex1_cml}-\cite{ex4_cml}.

In this work we study the dynamics of a collection of nonlinearly
coupled oscillators, relevant to systems ranging from neuronal
populations to electrical circuits, under coupling topologies varying
from a regular ring to a random network. We find that the trajectories
of this system escape to infinity under regular coupling, for
sufficiently strong coupling strengths, i.e. the amplitude of the
oscillators grow explosively without bound. We term this phenomenon a
{\em blow-up}. The surprising central result of this work is the
following: when some fraction of the regular connections in this
system are dynamically randomized, this unbounded growth is
suppressed.  Namely, {\em random links successfully reign in the
  blow-ups in this complex dynamical network}.

This counter-intuitive result is significant in the general context of
taming infinities, as infinities arising in theoretical constructs of
physical phenomena deeply concern scientists across disciplines
\cite{popular}. When physical quantities tend to infinite values in a
model, it is often considered an indication of the failure of a theory,
or symptoms of the limitations of the modelling. Removing infinities,
for instance through theoretical tools such as renormalization, or
through improvement of the underlying model of the physical process,
has had much impact on physics in general. Further, from the point of
view of potential applications, it is indeed crucial that no quantity
grows without bound. So the drive to find mechanisms that can quell
infinities in measurable quantities, stems from both methodological and
pragmatic motivations \cite{popular}. Specifically then, our
observations in this work may yield methods to control and prevent
blow-ups in extended oscillator systems, which are widely employed in
a multitude of engineered devices.
Lastly note that a certain degree of randomness in spatial links may
be closer to physical reality, in many systems of biological,
technological and physical significance \cite{Watts}. So our finding
that random links prevent un-physical blow-ups, suggests an underlying
mechanism by which complex systems can avoid explosive growths.\\
\newpage
{\bf Results}\\

We start by considering a generic coupled system comprised of
nonlinear local dynamics and a term modelling the coupling
interaction.  The isolated (uncoupled) dynamics at each node of the
lattice/network is given by $\dot{{\bf X}} = {\bf F}({\bf X})$,
where {\bf X} is a $m$-dimensional vector of dynamical variables and
{\bf F}({\bf X}) is the velocity field. A general form of a coupled
system of such dynamical elements is given by following equations:
\begin{equation}
  \dot{{\bf X}_{i}} = {\bf F}({\bf X}_{i}) + \epsilon \sum_{j=1}^{N}J_{ij}{\bf H} ({\bf X}_{i}, {\bf X}_{j}),\hspace{0.4in} i=1,..., N,
\label{general}
\end{equation}
where $J_{ij}$ are the elements of a connectivity matrix, $\epsilon$
is the coupling strength, and ${\bf H} ({\bf X}_{i},{\bf X}_{j})$ is a
coupling function determined by the nature of interactions between
dynamical elements \emph{i} and \emph{j} where $J_{ij}$ need not 
in general be symmetric matrix \cite{msf}.

At the nodes, we first consider the prototypical Van der Pol
oscillator \cite{vanderpol}, with nonlinear damping, governed by the
second-order differential equation:
$$\ddot{x}+\mu(x^2-1)\dot{x}+x=0$$
where $x$ is the dynamical variable and $\mu$ is a parameter
determining the nature of the dynamics. 
Associating $\dot{x}=y$ gives:
\begin{eqnarray} 
\dot{x}&=&f(x,y) = y \nonumber \\
\dot{y}&=&g(x,y) = -\mu(x^2-1)y-x
\label{vdp} 
\end{eqnarray}
The relaxation oscillations arising in this system, for $\mu > 0$, are
characterised by slow asymptotic behavior and sudden discontinuous
jumps. The Van der Pol equation above is very relevant in modelling
phenomena in physical and biological sciences. Extensions of such
oscillators have been used to model the electrical activity of the
heart and the action potentials of neurons \cite{vdp_app1}. The
equation has also been utilised in seismology to model the two plates
in a geological fault \cite{vdp_app2}.


We now consider a {\em nonlinearly coupled} ring of $N$ such
Van-der-Pol oscillators \cite{nlin_coup_1}-\cite{nlin_coup_blow_up},
namely a specific form of Eqn. \ref{general} (with $m=2$, ${\bf X}_i =
\{ x_i, y_i \}$, ${\bf F} = \{ f, g \}$ and ${\bf H} ({\bf X}_i, {\bf
  X}_j) = \{ f(x_j, y_j) - f(x_i,y_i), g(x_j,y_j) - g(x_i,y_i) \}$ )
given as follows:

\begin{eqnarray} 
{\dot{x}}_{i}&=&f(x_{i},y_{i}) + \frac{\varepsilon}{2} [ f(x_{i-1},y_{i-1}) + f(x_{i+1},y_{i+1}) - 2 f(x_i,y_i) ]  \nonumber \\
{\dot{y}}_{i}&=&g(x_{i},y_{i}) + 
\frac{\varepsilon}{2} [ g(x_{i-1},y_{i-1}) + g(x_{i+1},y_{i+1}) -2 g(x_i,y_i) ]
\label{array}
\end{eqnarray}

Here index $i$ specifies the site/node in the ring, with the nodal
on-site dynamics being a Van der Pol relaxation oscillator. 
Starting with this regular ring, we construct increasingly random
networks, with fraction of random links $p$ ($0 \le p \le 1$), as
specified in the methods section.

Additionally, we allow our links to {\em dynamically switch} at
varying rates, i.e. our underlying web of connections can change over
time \cite{dynamic_links1,dynamic_links2}. We implement this by
rewiring the network at time intervals of $r$. So a new adjacency
matrix, having the same fraction of random connections, is formed
after a time period of $r$. Such dynamic rewiring is expected to be
widely prevalent, for instance
in response to environmental influences or internal adaptations
\cite{dynamic1,dynamic2}. The parameter $r$ gives the
time-scale of network switches, and as $r$ becomes larger the
connections get more static, with the limit of $r \rightarrow \infty$
corresponding to the standard scenario of quenched random links.

We solved the $2N$ coupled non-linear ordinary differential equations
(ODE) given by Eqns.~\ref{vdp},\ref{array} in order to get the dynamics of the
complete system, under varying connection topologies and coupling
strengths, for a large range of system sizes. We study the shapes and
sizes of the limit cycles arising in networks with varying fractions
of random links $p$, and network switching time periods $r$.

We first describe our principal observations for oscillators coupled
to two nearest neighbours on a periodic lattice (i.e. $p = 0$) under
increasing coupling strengths.  When the strength of coupling is in
the range between $0$ and $\varepsilon_c$, the system yields regular
bounded behaviour.
However, when the coupling strength exceeds a critical value,
i.e. $\varepsilon > \varepsilon_{c}$, there is a {\em blow-up} in the
system.  Namely, the amplitude of the oscillations grows explosively
in an unbounded fashion
\cite{nlin_coup_blow_up},\cite{blowup_example}-\cite{ex2}. For
instance, in a typical blow-up, the magnitude of the amplitude grows
from $O(1)$ to $O(10^4)$ in time $\sim 10^{-3}$.



Now, very different behaviour from that described above, ensues when
the links are rewired randomly. As mentioned before, we consider
dynamically changing links, and introduce a time scale for the random
rewiring. The time period for changing links in the network is
$r$. Namely, a new adjacency matrix is formed, with the same fraction
of random links ($p > 0$), at time intervals of $r$. 

\begin{figure}[htb]
  \includegraphics[width=0.7\linewidth,angle=270]{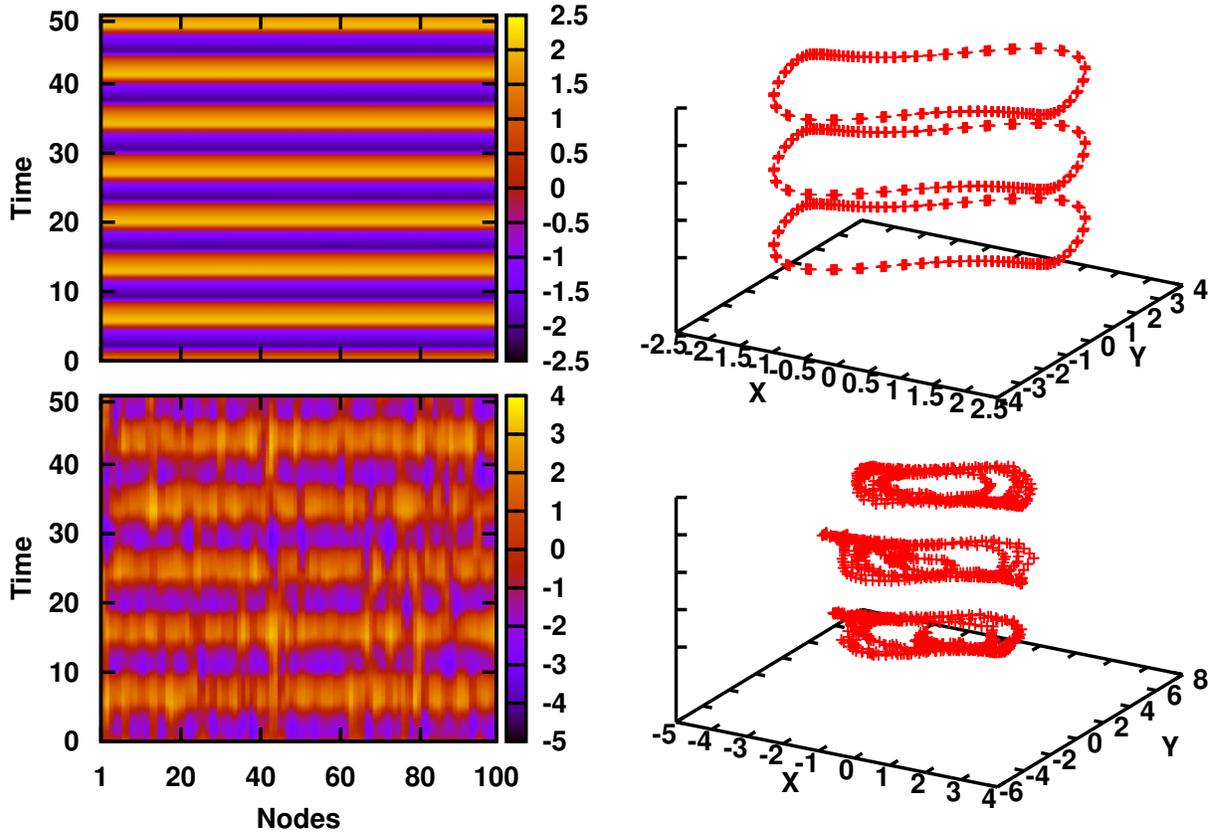}
     \caption{Evolution of $x_i (t)$ of all the oscillators ($i =1,
       \dots 100$), after a transient time of $15000$ (left), and the
       limit cycles of three representative oscillators stacked one
       above the other (right), for network rewiring period $r$ equal
       to $0.01$ (top row) and $0.1$ (bottom row). Here coupling
       strength $\varepsilon = 0.6$ (which is significantly greater
       than $\varepsilon_c$) and the fraction of random links $p$ is
       $0.8$. Blow-up is prevented in both the cases, but faster
       rewiring leads to a synchronized state (top row) whereas slower
       rewiring leads to de-synchronization and distortion in the
       shape of the limit cycles (right panel, bottom row). Note that
       for this value of coupling strength, regular nearest neighbour
       coupling ($p=0$) would lead to a blow-up.}
\label{rewired2}
 \end{figure}

We show our results for representative values of $p$ and $r$ in Figure
\ref{rewired2}. Notice that the coupling strengths shown in the figure
are {\em greater than the critical value} $\epsilon_c$, i.e. for these
coupling strengths the limit cycles grow explosively for regular
nearest neighbour coupling. However, it is clearly evident from the
figure that the {\em blow-up has been tamed} when some fraction of the
regular links are replaced by random links (i.e $p>0$). Namely, for
sufficiently random networks, all oscillators remain confined in a
bounded region of phase space.

\begin{figure}[htb]
\begin{center}
 \includegraphics[width=0.7\linewidth,angle=270]{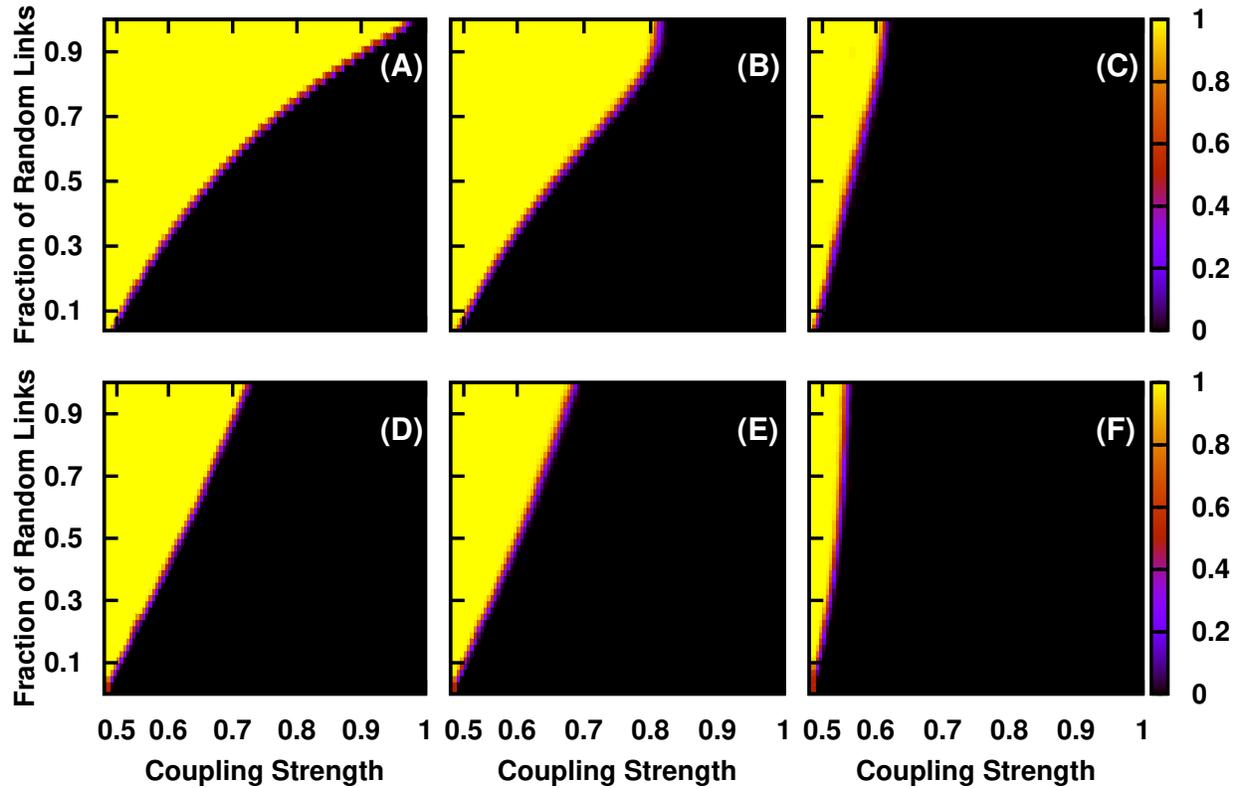}
\end{center}
\caption{Transition between bounded and unbounded dynamics in a system
  of $100$ coupled oscillators, as reflected in the variation of the
  boundedness order parameter $Z_{bound}$ (see text) with increasing
  coupling strength $\varepsilon$ and fraction of random links $p$,
  for directed networks with network switching time period $r$ equal
  to (A) $0.01$, (B) $0.1$ and (C) $1$, and for undirected networks
  with $r$ equal to (D) $0.01$, (E) $0.1$ and (F) $1$.}
\label{zbound}
\end{figure}

In order to characterize the bounded-unbounded transitions
quantitatively, we introduce a ``boundedness order parameter'',
$Z_{bound}$, which is given simply as the fraction of initial
conditions for which the dynamics remained bounded. Thus $Z_{bound}$
gives the probability of successful prevention of
blow-ups. Fig. \ref{zbound} displays this quantity over a range of $p$
and $\varepsilon$. It is evident from Fig. \ref{zbound}, that for
fraction of random links $p$ greater than a critical value, we always
get a bounded state.

The fraction of random links necessary to obtain bounded dynamics with
probability close to $1$ depends on the network switching time period
$r$ and coupling strength $\varepsilon$ as indicated in
Fig. \ref{zbound}. In general, for fast switching links, namely small
$r$, the critical number of random links needed to suppress a blow-up
is small. For very slow network changes however, one may not be able
to quell explosive growth even with a large fraction of random
links. So it is clear that {\em rapidly changing links} is the key to
controlling the dynamics more efficiently.

Further, we studied the case of {\em undirected networks} (where $J$
in Eqn.~\ref{general} is a symmetric matrix).  Qualitatively, similar
results are obtained, though with reduced range for blow-up control
(see lower panel of Fig. \ref{zbound}).

Now, in order to broadly rationalize this behavior we consider the
dynamics of an individual oscillator in the system as a sum of the
nodal dynamics and the coupling contributions of the form $c_1(t)$ and
$c_2(t)$, whose strength is given by coupling constant $\epsilon$.
\begin{eqnarray}
\dot{x}&=&y + \epsilon c_{1}(t) \nonumber \\
\dot{y}&=&-\mu(x^2-1) y - x +  \epsilon c_{2}(t)
\label{c1_c2}
\end{eqnarray}
When $c_{1}=0$ and $c_{2}=0$, the equations reduce to a single Van
der-Pol oscillator, with a stable limit cycle and an unstable spiral
repellor at the origin ($x^* = 0, y^*=0$). When $c_1 \neq 0$, $c_2
\neq 0$, the effective dynamics of each oscillator in the network
(Eqn.~\ref{c1_c2}) can be seen as an oscillator with fluctuating
parameters $c_1$ and $c_2$ \cite{fuzzy}. Note that when $c_1 \neq 0$
one obtains the fixed point solutions to be:
\begin{eqnarray}
x^*&=&\frac{1 \pm \sqrt{1- 4 \mu  \epsilon^2 c_1 (c_{2}-\mu c_{1})}}{2 \epsilon \mu c_{1}} \nonumber \\
y^*&=& - \epsilon c_{1}
\end{eqnarray}
This implies that we get two fixed point solutions for each oscillator
in the coupled system. For very weak coupling ($\varepsilon
\rightarrow 0$) the additional fixed point sits at infinity. As
coupling strength increases this fixed point migrates towards the
limit cycle, and at some critical coupling strength collides with it,
undergoing a saddle node bifurcation. This global bifurcation leads to
the destruction of the attracting limit cycle and consequent
blow-up. Now for noisy parameters $c_1$ and $c_2$, such a collision
leading to an annihilation of the limit cycle occurs for sufficiently
large fluctuations in $c_1$ and $c_2$ \cite{fuzzy}. Since the
magnitude of fluctuations is governed directly by coupling strength,
this implies that blow-ups may occur for sufficiently large
$\varepsilon$.

In this scenario, the underlying mechanism that allows the control of
blow-up for random links is the following: when the links are changed
rapidly (i.e. $r$ is small), to first approximation, the terms
$c_1(t)$ and $c_2(t)$ in Eqn.~\ref{c1_c2}, averages to zero, as it is
a sum of uncorrelated random inputs. This effectively yields the
condition for a single Van der Pol oscillator, namely no
blow-up. Alternately, one can argue that for rapid random switching of
coupling connections, the deviations from the single oscillator
dynamics, is essentially a sum of uncorrelated contributions, which to
first approximation average out to zero.\\


\bigskip

\noindent
{\bf Synchronization of the Bounded State}\\

Returning to the space-time plots displayed in Fig.\ref{rewired2} we
notice an additional feature. In the figure two cases with different
network rewiring time periods are shown: a fast rewiring case and the
slow network change scenario.  While there is no blow-up in either
case, the emergent bounded state is different for networks rewiring at
different rates. When the underlying network changes rapidly one
obtains a synchronized state \cite{barahona}-\cite{hong}, as evident
in the representative example in the upper panel of
Fig. \ref{rewired2}. However, slow network changes yield bounded
dynamics that is not synchronized, as typically seen in the
representative example in the lower panel of Fig. \ref{rewired2}.

So we will now investigate, in detail, the synchronization properties
of the system. Specifically we attempt to answer the following
question: when random coupling prevents a blow-up, does it yield a
state that is synchronized, or not? In order to quantify the degree of
synchronization in the network, we compute an average error function
as the synchronization order parameter, defined as the mean square
deviation of the instantaneous states of the nodes:
\begin{equation}
Z =\frac{1}{N}\sum_{i=1}^N\{[x_i (t) - \langle x (t) \rangle]^2\},
\label{eq:defz}
\end{equation}
where, $\langle x (t)\rangle$ is the space average at time $t$. This
quantity $Z$, averaged over time $n$ and over different initial
conditions, is denoted as $Z_{sync}$. When $Z_{sync}=0$, we have
complete synchronization.

Note that the time needed to reach synchronization is not simply
monotonic with respect to fraction of random links
\cite{intermediatep}. Instead we observe a region, at intermediate
values of $p$, where the transience becomes large. This kind of
behavior is reminiscent of the long persistence of transience arising
for intermediate values of network connectivity \cite{transience}. So,
in our calculation of the relevant order parameters we have left
sufficiently long transience ($\sim 15000$).

\begin{figure}[htb]
  \includegraphics[width=0.75\linewidth,angle=270]{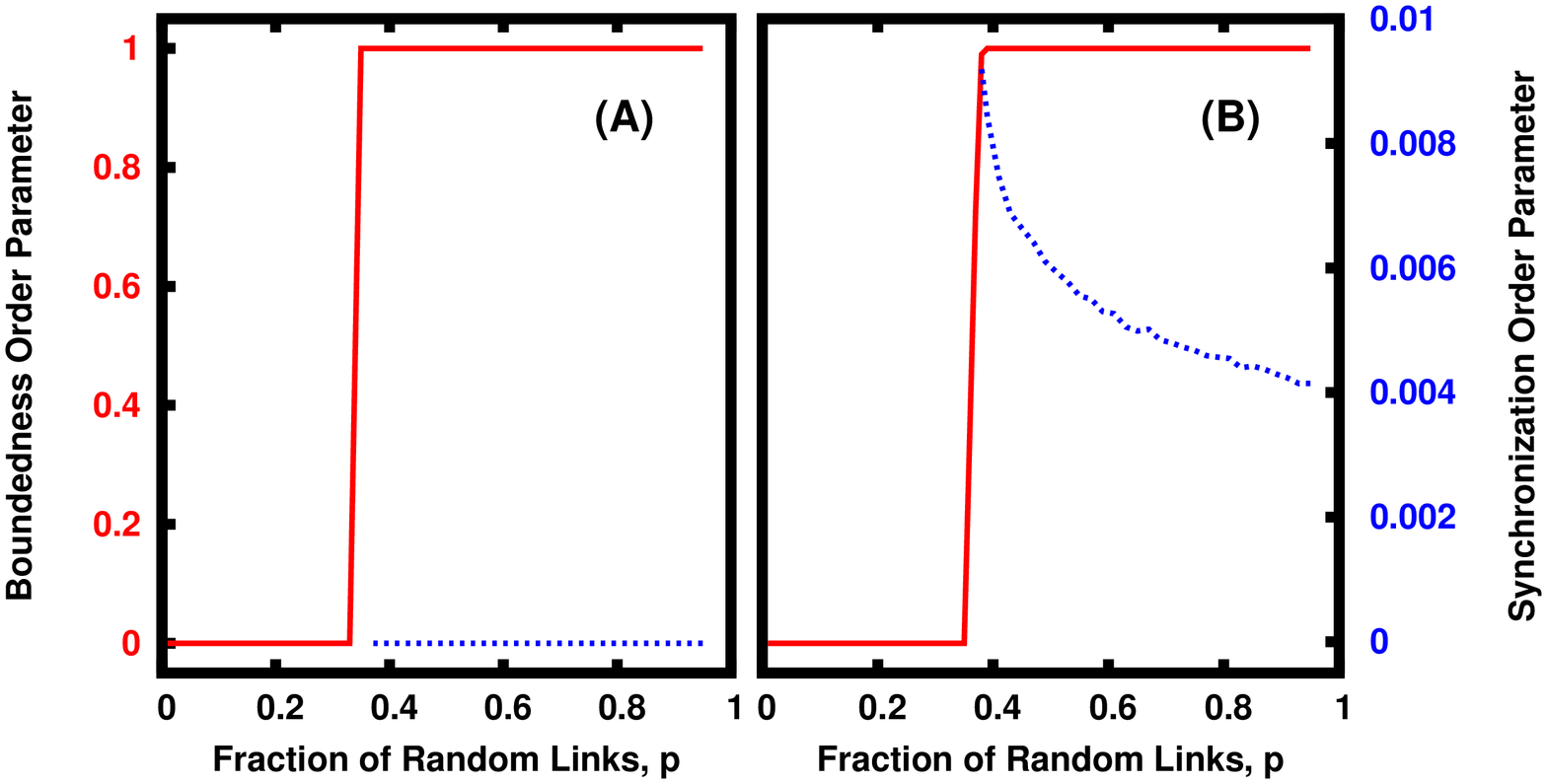}
 \caption{Variation of the Boundedness Order Parameter $Z_{bound}$
   (solid red curve) and the Synchronization Order Parameter
   $Z_{sync}$ (dotted blue curve), averaged over $100$ initial
   conditions, with respect to fraction of random links $p$ for
   network rewiring time period (A) $r = 0.01$ and (B) $r = 0.1$. Here
   system size $N=100$ and coupling strength $\epsilon = 0.6$, which
   is significantly greater than $\varepsilon_c$.}
\label{zsync}
\end{figure}

It is evident from Fig. \ref{zsync} that in this particular system,
for $p=0$, namely a regular ring, there is {\em no synchronization},
even at the largest allowed coupling strengths before blow-up, i.e. no
$\epsilon < \epsilon_{c}$ yields a synchronized state. However for
networks that change connections fast (i.e. $r$ is sufficiently small)
the bounded state that arises after random rewiring is also
synchronized. This is evident, quantitatively, through the vanishingly
small synchronization error ($Z_{sync} \sim 0$) in networks where
blow-ups have been completely suppressed (i.e. $Z_{bound} = 1$).  
On the other hand, for slowly changing networks, the synchronization
error is significant, even when the dynamics is bounded,
i.e. $Z_{sync}$ is not close to zero even when $Z_{bound} = 1$, as
evident in Fig. \ref{zsync}b. So infrequent network changes have led
to a state which is {\em bounded but not synchronized}, i.e. slowly
switching random links are effective in reigning in the unbounded
dynamics, but they do not manage to produce synchronization.

This relaxation oscillator network then yields three kinds of
dynamical states: (i) bounded synchronized motion; (ii) bounded
un-synchronized dynamics; and (iii) unbounded dynamics. The nature of
the dynamics depends on the interplay of the coupling strength
$\varepsilon$, rewiring probability $p$ and network switching time
$r$.\\

\bigskip

\noindent
{\bf Probabilistic Switching of Links}\\

Now, periodic switching of links occurs in situations where the
connections are determined by a global external periodic
influence. However this is not always the most realistic scenario, as
the interaction patterns may not change periodically in time. In such
cases we must consider a probabilistic model of link switching, such
as in \cite{Vazquez2010_1,Vazquez2010_2}. So in this subsection we
study such randomly switched networks in order to determine if the
ability to tame explosive growth is robust to the manner in which the
links change.  Namely, we verify the generality of our observations
above by investigating the dynamics on a network whose underlying
links switch randomly and asynchronously in time.

Specifically now, at each instant of time, a link has a probability
$p_r$ of being rewired (the ``link rewiring probability''). So when
$p_r$ is higher the links are switched around more often. Note that as
before, when a node rewires, it links to its nearest neighbours with
probability $1-p$ and to some random site with probability $p$.

\begin{figure}[htb]
\begin{center}
   \includegraphics[width=0.7\linewidth,angle=270]{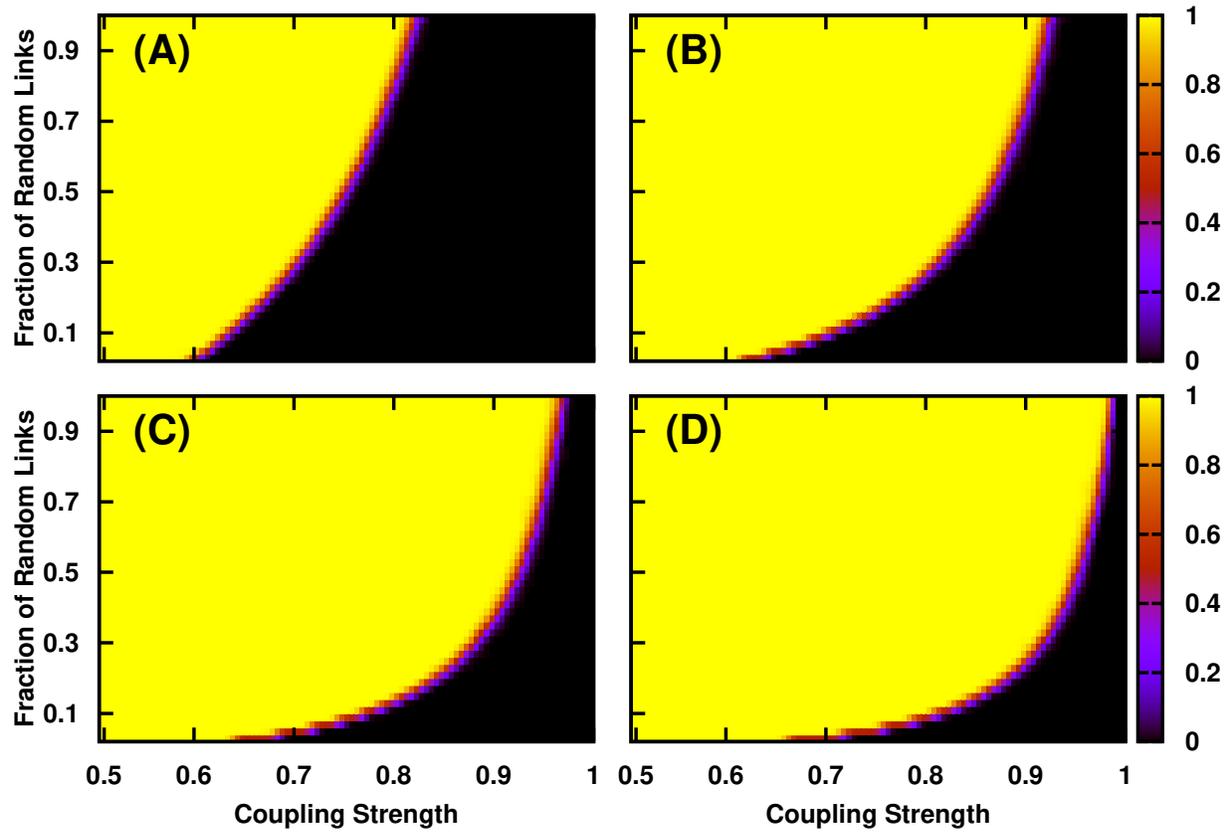}
\end{center}
\caption{Transition between bounded and unbounded dynamics in a system
  of $100$ coupled oscillators with probabilistically switched links,
  as reflected in the variation of the boundedness order parameter
  $Z_{bound}$ (see text) with increasing coupling strength
  $\varepsilon$ and fraction of random links $p$. The link switching
  probability $p_r$ is: (A) $0.1$, (B) $0.3$ (C) $0.6$ and (D) $0.9$.}
\label{zbound_pr}
\end{figure}

\begin{figure}[htb]
 \includegraphics[width=0.55\linewidth,angle=270]{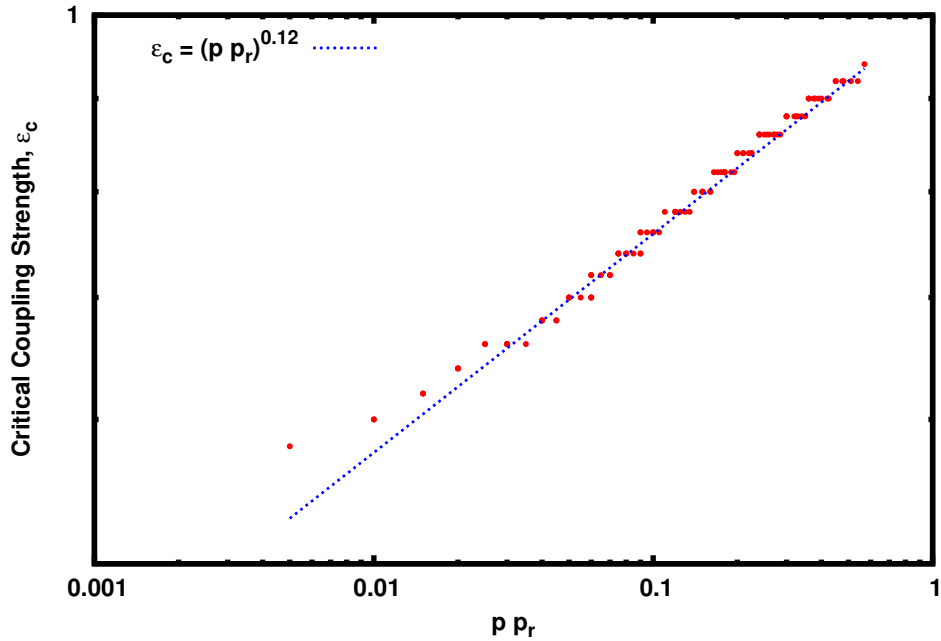} 
 \caption{Variation of the critical coupling strength at which blow-up
   occurs, $\varepsilon_c$, with respect to the product of the link
   rewiring probability $p_r$ and fraction of random links $p$. The
   points are data from a wide range of $p$ and $p_r$, and they
   collapse to the scaling form shown by the line: $\epsilon_c \sim (
   p \ p_r )^{\beta}$ with $\beta = 0.12$.}
\label{pr1}
\end{figure}

\begin{figure}[htb]
 \includegraphics[width=0.55\linewidth,angle=270]{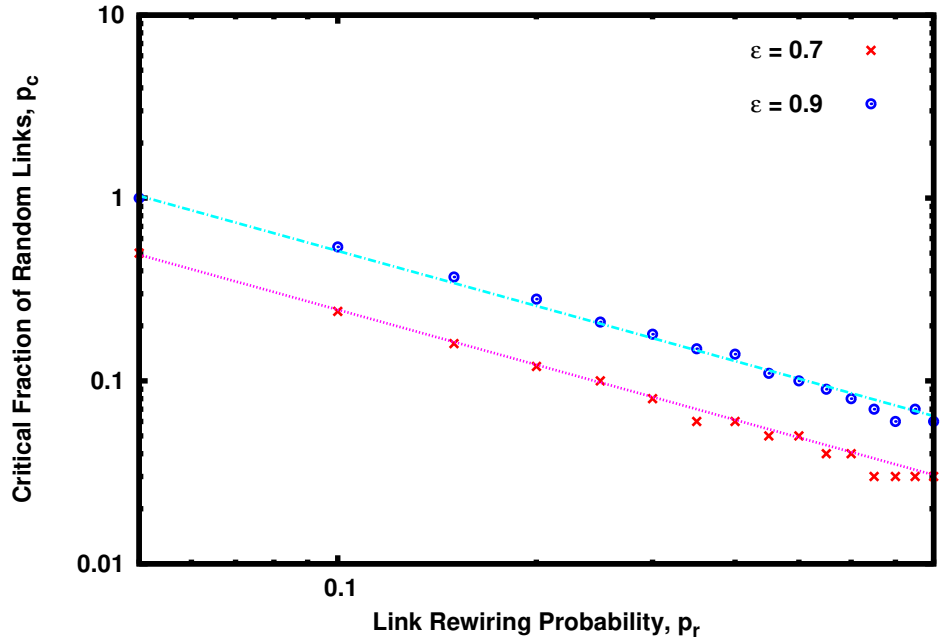} 
 \caption{Variation of the minimum fraction of random links necessary
   to prevent blow-ups vs link rewiring probability $p_r$, at coupling
   strengths $\epsilon = 0.7$ (red) and $0.9$ (blue). The straight
   lines show the fit to the scaling form: $p_c \sim p_r^{-1} $}
\label{pr2}
\end{figure}

\begin{figure}[htb]
    \includegraphics[width=0.7\linewidth,angle=270]{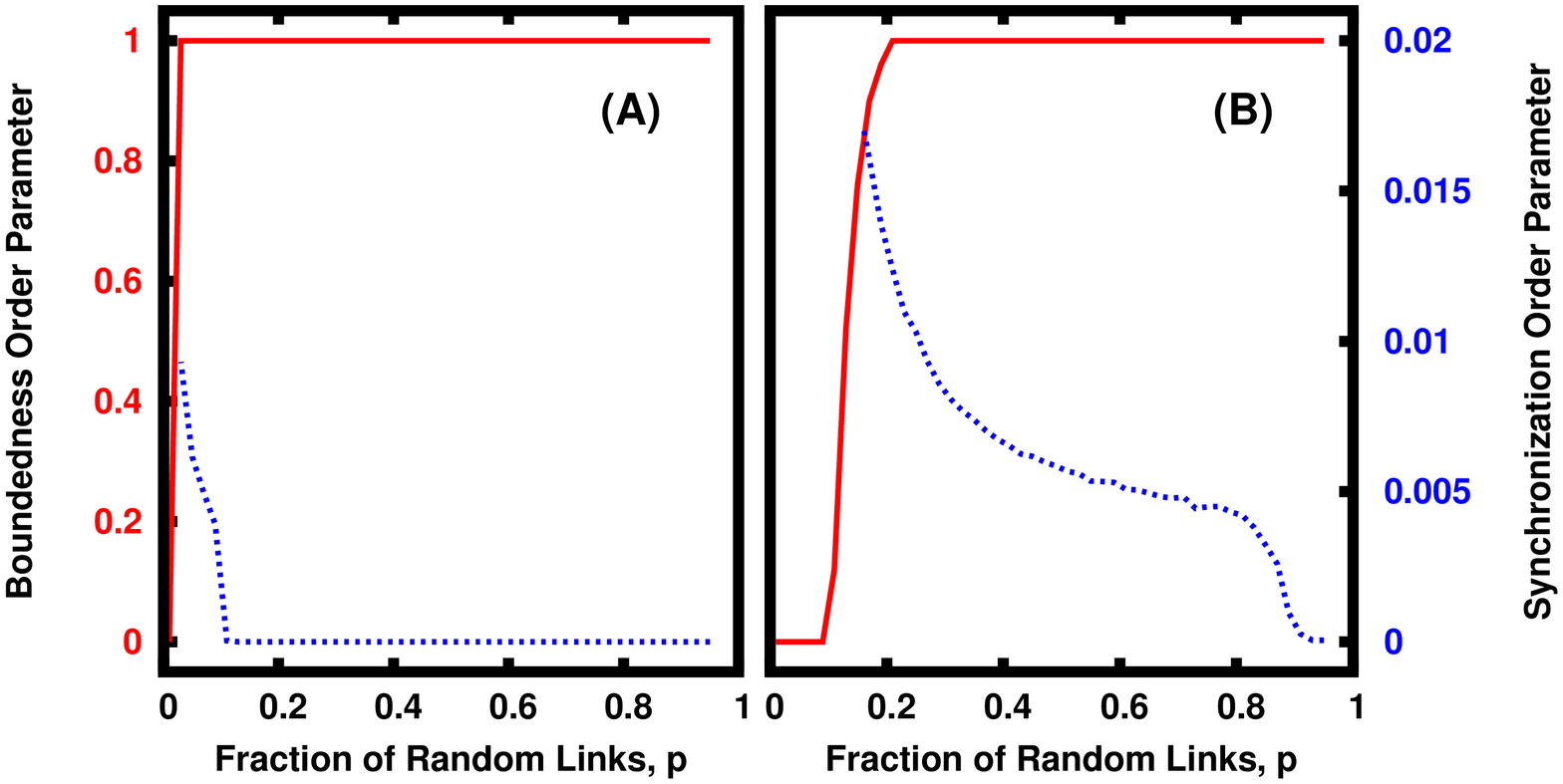}
 \caption{Variation of the Boundedness Order Parameter $Z_{bound}$
   (solid red curve) and the Synchronization Order Parameter
   $Z_{sync}$ (dotted blue curve), averaged over $100$ initial
   conditions, with respect to fraction of random links $p$ for
   link rewiring probability (A) $p_r = 0.9$ and (B) $p_r = 0.1$. Here
   system size $N=100$ and coupling strength $\epsilon = 0.6$, which
   is significantly greater than $\varepsilon_c$.}
\label{zsync_pr}
\end{figure}

The results are displayed in Figs. \ref{zbound_pr}-\ref{zsync_pr}.  As
in the case of periodically changing networks, here too random links
manage to control the explosive growth in the system, for sufficiently
large fraction of random links. Also, clearly, when the probability of
rewiring the links is larger, namely when the network changes more
often, the critical coupling strength at which blow-up occurs is
pushed back, yielding a larger region of bounded dynamics.

Further, we observe that for networks that change connections fast
(i.e. when $p_r$ is large) the bounded state that arises after random
rewiring is also {\em synchronized}, as evident quantitatively through
the vanishingly small synchronization error in these networks
(cf. Fig. \ref{zsync_pr}a). However, for slowly changing networks
(i.e. when $p_r$ is small), the synchronization error is quite
significant, even when the dynamics is bounded
(cf. Fig. \ref{zsync_pr}b), namely, slowly changing links yield a
state that is {\em bounded but not  synchronized}.

Inspecting further the interplay of the fraction of random links $p$,
and the link rewiring probability $p_r$, reveals some interesting
scaling relations. Notice first that the critical coupling strength
$\epsilon_c$ beyond which blow-ups occur, scales with the link
rewiring probability $p_r$ and fraction of random link $p$ as:
$$\epsilon_c \ \ \sim \ \  ( p \  p_r )^{\beta}$$
with $\beta \sim 0.12$. This is clearly evident, quantitatively, in the
data collapse displayed in Fig. \ref{pr1}. This scaling relation
implies that as the connections change more frequently, and there is
larger fraction of random links in the connectivity matrix, the range
over which bounded dynamics is obtained becomes larger. Also notice
that the {\em effective} quantity that is relevant in the scaling
relation is the product of $p$ and $p_r$, i.e. $p$ and $p_r$ may vary,
but the emergent phenomena is the same if their product remains the
same.

Further calculating the minimum fraction of random links $p_c$
necessary to prevent blow-ups, for a fixed coupling strength, at
different link rewiring probabilities $p_r$ (cf. Fig. \ref{pr2})
reveals that
$$p_c \ \ \sim \ \ p_r^{-1} $$
Namely, when the system has large spatial randomness, infrequent
changes of links manages to suppress blow-ups, while in a system with
few spatially random connections, the links need to change more often
in order to effect containment of blow-ups. This again implies (as did
the scaling form in Fig. \ref{pr1}) that blow-ups can actually be
tamed at quite small link rewiring probabilities and in networks with
small fraction of random links, underscoring the broader relevance of
our
observations.\\

\bigskip

\noindent
{\bf Generality of our central result}\\

Finally, in order to further gauge the generality of the above
findings, we investigated different coupling forms and local dynamics.
For instance, we studied some other forms of nonlinear coupling, such
as:

\begin{eqnarray} 
{\dot{x}}_{i}&=&f(x_{i} + \frac{\varepsilon}{2} [ x_{i-1} + x_{i+1} - 2 x_i], y_{i} + \frac{\varepsilon}{2} [ y_{i-1} + y_{i+1} - 2 y_i])  \nonumber \\
{\dot{y}}_{i}&=&g(x_{i} + \frac{\varepsilon}{2} [ x_{i-1} + x_{i+1} - 2 x_i], y_{i} + \frac{\varepsilon}{2} [ y_{i-1} + y_{i+1} - 2 y_i])
\label{array1}
\end{eqnarray}

and

\begin{eqnarray} 
{\dot{x}}_{i}&=&f(x_{i},y_{i}) + \frac{\varepsilon}{2} [ x_{i-1}^{2} + x_{i+1}^{2} - 2 x_{i}^{2}]) \nonumber \\
{\dot{y}}_{i}&=&g(x_{i},y_{i}) + \frac{\varepsilon}{2} [ y_{i-1}^{2} + y_{i+1}^{2} - 2 y_{i}^{2}])
\label{array2}
\end{eqnarray}

We also investigated the case where only one dynamical equation is coupled:

\begin{eqnarray} 
{\dot{x}}_{i} &=& f(x_{i},y_{i}) + \frac{\varepsilon}{2} [ f(x_{i-1},y_{i-1}) + f(x_{i+1},y_{i+1}) - 2 f(x_i,y_i) ] \nonumber \\
{\dot{y}}_{i} &=& g(x_{i},y_{i})
\label{array3}
\end{eqnarray}
Since $f(x,y) = y$ in the Van der Pol oscillator, this coupling is
similar to linear coupling of conjugate or dissimilar variables. The
stability of the synchronized state in this system is amenable to
analysis through the master stability function formalism \cite{msf} and
shows the following: the maximum Floquet multiplier remains positive
for at least one eigen-mode, for all coupling strengths, implying that
the synchronized solution is not stable under this form of coupling in
regular networks.

Now, simulating all these different coupling forms above
(Eqns.~\ref{array1}, \ref{array2} and \ref{array3}), with some
fraction of links randomized, displayed the same qualitative
phenomena: After a certain critical coupling strength we encountered a
blow-up in a regular ring of these oscillators, with the blow-ups
occurring very early in some systems (e.g. $\epsilon_c < 0.1$ in
Eqn. \ref{array2}). However, as observed in the earlier systems,
dynamically randomizing the links in the network suppressed this
explosive growth yielding bounded dynamics over a larger range of
coupling strengths. Note that the range of the bounded region depends
on the specifics of the coupling form and the local dynamics, as well
as the fraction and frequency of change of the random links.

Lastly, we have also investigated the global behaviour of the system
under {\em different nodal dynamics}. For instance, we investigated
collections of nonlinearly coupled Stuart-Landau oscillators, where
the local dynamics was given by:
\begin{align}
\dot{z}=(1+i\omega-|z|^2)z
\end{align}
where $\omega$ is the frequency, and $z(t)=x(t)+iy(t)$.


We also studied nonlinearly coupled networks of FitzHugh-Nagumo
oscillators, modelling neuronal populations, described by the local
dynamics \cite{fn1,fn2}:
\begin{eqnarray}
 \dot{x}&=& x - x^{3}/3 - y + I \nonumber \\
 \dot{y}&=& \varPhi(x + a - by) 
\label{fn}
\end{eqnarray}
where {\em x} is the membrane potential, {\em y} is the recovery
variable and {\em I} is the magnitude of stimulus current. 

Further, we have explored the {\em heterogeneous} case where the
nonlinearity parameter $\mu$ is distributed over a range of positive
values for the local Van der Pol oscillators. 

\begin{figure}[htb]
    \includegraphics[width=0.7\linewidth,angle=270]{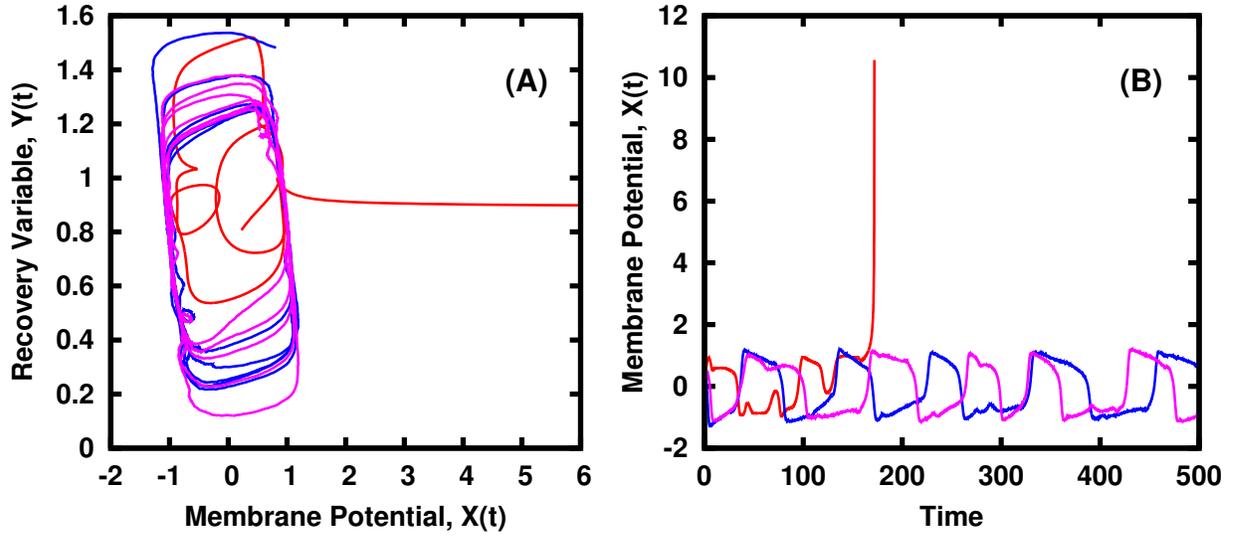}
\caption{ Phase portrait(left) and Time evolution of the membrane potential (right) for three representative FitzHugh-Nagumo model
  neurons in the network given by Eqn. 11, with fraction of
  random links $p = 0.5$, time period of changing the links $r =
  0.01$, and coupling strength $\epsilon = 0.54$ (cyan and blue
  curves). The case of a static regular network, exhibiting blow-up,
  is shown alongside as well (red curve). Note that the bounded orbits
  obtained under dynamic random links are not synchronized (cyan and
  blue curves).}
\label{FN}
 \end{figure}

Our extensive numerical simulations showed qualitatively similar
behaviour for all of the above systems (see Fig. \ref{FN} for
representative examples). This strongly suggests that the phenomenon
of prevention of blow-ups through dynamic random links
is quite general.\\

{\bf Conclusions}\\

In summary, we have studied the dynamics of a collection of
relaxation oscillators, under varying coupling topologies, ranging
from a regular ring to a random network. We find that the system blows
up under regular ring topology, for sufficiently strong coupling
strength. However, when some fraction of the links are randomized
dynamically, this blow-up is suppressed and the system remains
bounded. So our results suggest an underlying mechanism by which
complex systems can avoid a catastrophic blow-up. Further, from the
stand point of potential applications, our observations indicate a
method to control and prevent blow-ups in coupled oscillators that are
commonplace in a variety of engineered systems.\\

\bigskip

{\bf Methods}\\

\bigskip

We first present the method for changing the connectivity of the network.\\

{\em Algorithm for Periodically Switching Networks:}\\

\begin{enumerate}
\item We start with a $1$ dimensional ring, with $N$ nodes, where $N$ is the
system size. Each node $i$ ($i =1,\dots N$) has $2$ links. In the
network, a fraction $p$ of the links are to random nodes in the ring,
and fraction $(1-p)$ are to nearest neighbours.  So this network
corresponds to a standard small world network \cite{Watts} with degree
$2$, with fraction of random links given by $p$.

\item 


After time $r$, change the $2$ existing connections for all nodes
$i$. The new links are chosen such that, again, with probability $p$
they go to random neighbours, and with probability $(1-p)$ they are to
nearest neighbours. So the changed network is also a standard small
world network with degree $2$, with fraction of random links given by
$p$.  

\item Repeat step 2 again after every $r$ time steps.

\end{enumerate}

  So, while the fraction of random links is always $p$, the
  connectivity matrix of the system changes every $r$ steps. That is,
  $r$ is the network rewiring period.\\

{\em Algorithm for Stochastically Switching Networks:}\\

\begin{enumerate}
\item Again start with a $1$ dimensional ring, with $N$ nodes, where
  each node $i$ ($i =1,\dots N$) has $2$ links, and a fraction $p$ of
  the links are to random nodes, and fraction $(1-p)$ are to nearest
  neighbours. 

\item Now at every instant of time, a node $i$ has probability $p_r$ to
  change its links. That is, $p_r$ gives the link rewiring
  probability of each individual node.

\item If the $2$ links of a certain node are changing, the new links
  are chosen such that, again, with probability $p$ they connect to
  random neighbours, and with probability $(1-p)$ they connect to
  nearest neighbours.

\end{enumerate}

So in the above algorithm, the nodes change their links independently
and stochastically, with probability $p_r$. That is, at any given
time, fraction $p_r$ of the links change on an average, and $(1-p_r)$
retain their existing links. The probability of the new links being
random or regular is determined by $p$, as in the first algorithm (and
as in the standard small-world scenario.)

Please note that in the first algorithm $r$ is the {\em time period}
of link changes which occur globally throughout the network, while in
the second algorithm the change occurs at the local level with
{\em probability} $p_r$.

Our simulations show, that these two very different ways of varying the
network, both yield the same qualitative phenomena: namely changing
links helps to suppress blow-ups in the system.\\

{\em Method for simulating the dynamics of the system:} We solved the
$2N$ coupled non-linear ordinary differential equations (ODE), given
for instance by Eqns. \ref{array}, \ref{array1}, \ref{array2},
\ref{array3}, by integrating the ODEs with the fourth order
Runge-Kutta algorithm.  We have checked that the results are
qualitatively same over a large range of step sizes ($\sim 10^{-3}$ to
$10^{-7}$), in order to rule out artefacts arising from the stiffness
problem in such differential equations.

Using the above, we obtained the shapes and sizes of the limit cycles
arising in this network under varying fraction of random links $p$ ($0
\le p \le 1$) and a large range of network switching time periods $r$
($0.01 \le r \le 1$), for system sizes ranging from $N=10$ to
$N=10^3$.\\

{\em System Parameters :} In the system given by Eqns.~\ref{vdp} and
\ref{array}, we have investigated a range of nonlinearity parameters
$\mu$, and we present the representative case of $\mu = 1.5$ in the
examples displayed in this paper. In the system given by
Eqn.~\ref{fn}, parameters $\varPhi, a, b$ are taken to be $0.08, 0.7,
0.8$ respectively, in our
simulations.\\

\bigskip
\newpage
{\bf Acknowledgements}\\

AC and VK acknowledge research fellowship from CSIR, India.\\

\bigskip

{\bf Author contributions}\\

SS conceived the problem, and AC and VK performed all the numerical
simulations. SS, AC and VK discussed the results and wrote the
manuscript together.\\

\bigskip

The authors declare no competing financial interests.\\

\bigskip
\newpage

\end{document}